\documentclass[prl,preprint,showpacs,aps]{revtex4-1}
\usepackage{amssymb}
\usepackage{epsfig}
\usepackage{amsmath}
\usepackage{CJK}
\usepackage{supertabular}
\usepackage{textcomp}

\begin{document}

\author{Yong-Hui Xia}
\author{Chang Xu }\email{cxu@nju.edu.cn}
\author{Hong-Shi Zong}\email{zonghs@nju.edu.cn}
\address{Department of Physics, Nanjing University, Nanjing 210093, China}

\title{A new approach for calculating nuclear symmetry energy}
\begin{abstract}

By using the functional path integral method, we obtain a model-independent formula for nuclear symmetry energy, which explicitly shows the relation between nuclear symmetry energy and isospin susceptibility. The latter one is found to be a probe to the QCD chiral phase transition. We further found that, the nuclear symmetry energy has an abrupt change at the critical nuclear density where the chiral symmetry restores partially, which could be detected from the experiments.

\bigskip


\end{abstract}

\maketitle

Nuclear symmetry energy $E_{sym}(\rho_B)$ plays an important role in understanding many issues in both nuclear physics and astrophysics \cite{Li:2001pet,Brown:2000pd,Danielewicz:2002pu,Baran:2004ih,Li:2008gp,Sharma:2010db,MYERS19661,Moller:1993ed,Pomorski:2003mv}. However, our current knowledge of $E_{sym}(\rho_B)$ is still poor, especially at supra-saturation densities \cite{Tsang:2008fd,Ou:2015jan,Hen:2014nza}. As well known, the origin of nuclear symmetry energy comes from both the Pauli exclusion principle and the isospin dependence of strong force. In the quasi-particle model, the symmetry energy can be decomposed into a {\it Pauli exclusion contribution} and a {\it symmetry potential contribution} \cite{Xu:2009bb,Xu:2010fh,Xu:2010kf}. The symmetry energy can also be computed by using sophisticated theoretical models with different force parameters \cite{Moller:2012pxr,Ropke:2012qv,Vidana:2011ap,Carbone:2011wk,Rios:2013zqa}. Currently, the predictions of different model calculations are diverse at supra-saturation densities (see Ref. \cite{Li:2008gp} for a review) and show a strong model dependence. In this work, we give a model-independent way to formulate symmetry energy using the functional path integral method, which is considered to be guided by the first principle. More importantly, we show that the nuclear symmetry energy has an abrupt change at the critical nuclear density where the chiral phase transition undergoes.

We firstly start from the partition function that is the crux of statistical mechanics. Once the partition function is known, all the thermal observables can be easily obtained. In order to obtain the thermal properties of  nuclear matter, we try to give a direct way to calculate the partition function within the approach of functional path integral. The partition function of nuclear matter ( proton and neutron ) at finite chemical potential and zero temperature in the Euclidean space is given by
 \begin{eqnarray}
\mathcal{Z}&=&\int \mathcal{D}\bar \Psi_{n}\mathcal{D} \Psi_{n}\mathcal{D}\bar \Psi_{p}\mathcal{D} \Psi_{p}\mbox{exp}[-S_{eff}+\int d^4 x \mu_{n}\bar \Psi_{n} \gamma_4 \Psi_{n}+\int d^4 x \mu_{p}\bar \Psi_{p} \gamma_4 \Psi_{p}] \nonumber \\
&=&\int \mathcal{D}\bar \Psi_{n}\mathcal{D} \Psi_{n}\mathcal{D}\bar \Psi_{p}\mathcal{D} \Psi_{p}\mbox{exp}[-S_{eff}+\int d^4 x \mu_{B}(\bar \Psi_{n} \gamma_4 \Psi_{n}+\bar \Psi_{p} \gamma_4 \Psi_{p})\nonumber\\
&&+\int d^4 x \mu_{I}(\bar \Psi_{n} \gamma_4 \Psi_{n}-\bar \Psi_{p} \gamma_4 \Psi_{p})],
\end{eqnarray}
where $S_{eff}$ is the effective nucleon field action. $\Psi_{n}$ and $\Psi_{p}$ denote the nucleon fields. $\mu_n$ and $\mu_p$ denote proton and neutron chemical potentials, respectively. $\mu_B = \frac{\mu_n+\mu_p}{2}$ is the baryon chemical potential and $\mu_I=\frac{\mu_n-\mu_p}{2}$ is the isospin chemical potential. The pressure of nuclear matter is given by
 \begin{eqnarray}
 P=\frac{\mbox{log}\mathcal{Z}}{\mathcal{V}},
 \end{eqnarray}
 where $\mathcal{V}$ is the four--dimension volume. According to method proposed in Ref. \cite{Zong:2008sm}, one can immediately obtain the total differential of pressure
\begin{eqnarray}
\mbox{d}P=\rho_B \mbox{d} \mu_B +\rho_I \mbox{d} \mu_I,\label{p}
\end{eqnarray}
where $\rho_B=\rho_n+\rho_p$ denotes the nucleon number density and $\rho_I=\rho_n-\rho_p$ is the asymmetry density denoting the difference between neutron and proton number densities. According to the thermal mechanics relation, the nuclear matter energy density is given by
\begin{eqnarray}
\varepsilon=-P +\rho_B \mu_B +\rho_I \mu_I.
\end{eqnarray}
Then the total differential of energy density is obtained as follows
\begin{eqnarray}
\mbox{d} \varepsilon =\mu_B \mbox{d}\rho_B+\mu_I \mbox{d} \rho_I. \label{E}
\end{eqnarray}
Integrating both sides of  Eq. (\ref{E}), one can obtain the energy per nucleon $E(\rho_{B},\rho_{I})$
\begin{eqnarray}
E(\rho_{B},\rho_I) &=& \frac{\varepsilon(\rho_B,\rho_I)-\varepsilon(\rho_B=0,\rho_I=0)}{\rho_B}\nonumber \\
&=&\frac{\int ^{\rho_B}_0\mu_B (\rho^{'}_B,\rho_I=0)d \rho_B^{'}}{\rho_B}+\frac{\int ^{\rho_I}_0\mu_I(\rho_B,\rho^{'}_I)d \rho_{I}^{'}}{\rho_B}.\label{Ep}
\end{eqnarray}
Thus, the energy per nucleon $E(\rho_{B},\rho_I)$ is split into two terms.  The first term stands for the symmetric part contribution and the second term stands for the asymmetric part contribution. This result is model-independent and valid even at $\rho_I=\pm\rho_B$ (for example, the neutron star case). Obviously, all the information of symmetry energy can be extracted from the second term. By performing the Taylor expansion on the second term in Eq. (\ref{Ep}) at $\rho_I=0$, then energy per nucleon becomes
\begin{eqnarray}
E(\rho_{B},\rho_I)=E_{0}(\rho_B,\rho_I=0)+\mu_I(\rho_B,\rho_I=0)\frac{\rho_I}{\rho_B}+\frac{1}{2}\rho_B (\frac{\partial \rho_I}{\partial\mu_I}\Big |_{\rho_{I}=0})^{-1} (\frac{\rho_I}{\rho_B})^{2}+O((\frac{\rho_{I}}{\rho_B})^3),\label{Eexp}
\end{eqnarray}
where $E_{0}(\rho_B,\rho_I=0)$ denotes energy per nucleon  in symmetric nuclear matter.  If the $SU(2)$ symmetry is taken into consideration, the $Z(2)$ group (exchanging symmetry between proton and neutron) as a subgroup of $SU(2)$ group is kept naturally, then all the odd powers terms in Eq. (\ref{Eexp}) are zero.   Now,  the symmetry energy is obtained as
\begin{eqnarray}
E_{sym}(\rho_B)=\frac{1}{2}\rho_B (\frac{\partial \rho_I}{\partial \mu_I}\Big|_{\rho_{I}=0})^{-1}, \label{sym}
\end{eqnarray}
where $\frac{\partial \rho_I}{\partial \mu_I}$ denotes the isospin susceptibility. Here, it should be stressed that the nuclear symmetry energy formula Eq. (\ref{sym}) is model-independent. The nucleon number density $\rho_B$ and asymmetry density  $\rho_I$ can be written in the forms as follows (more details can be found in Ref. \cite{Zong:2008zzb})
\begin{eqnarray}
\rho_B=-\mbox{Tr}[\gamma_4 G_n(\mu_B,\mu_I)+\gamma_4 G_p(\mu_B,\mu_I)],\nonumber\\ \label{nG}
\rho_I=-\mbox{Tr}[\gamma_4 G_n(\mu_B,\mu_I)-\gamma_4 G_p(\mu_B,\mu_I)],
\end{eqnarray}
 where $\mbox{Tr}$ denotes the trace of Dirac spinor and the integral of 4--dimension momentum. $G_p(\mu_B,\mu_I)$ and $G_n(\mu_B,\mu_I)$ are dressed propagators of proton and neutron at finite baryon chemical potential respectively. According to the Lorentz structure analysis, the nucleon propagator has the general form in the Euclidean space as follows
 \begin{eqnarray}
 &&G_f(\mu_B,\mu_I)\nonumber \\
 &&=\frac{1}{i A({\vec{p}}^2,\mu_B,\mu_I,p_4) \gamma_4 p_4+iB({\vec{p}}^2,\mu_B,\mu_I,p_4)\gamma \vec{p} +C({\vec{p}}^2,\mu_B,\mu_I,p_4)\mu_f\gamma_4+D({\vec{p}}^2,\mu_B,\mu_I,p_4)},\nonumber\\ \label{G}
 \end{eqnarray}
 where $f = p,n$. $G_f(\mu_B,\mu_I)$ is the two-point  Green function that is considered as the simplest Green function in quantum field theory. From Eqs. [8-10], it is clearly shown that the nuclear symmetry energy is only related to the two-point Green function $G_f(\mu_B,\mu_I)$. Once the fine nucleon propagator is known (for instance, from the state-of-art lattice simulations), the symmetry energy can be computed at all densities. Currently, it is still difficult to calculate Eq. (\ref{G}) from the first principle, one has to resort some theoretical models, for example, the quasi-particle approximation. In the quasi-particle approximation, our present result Eq. (\ref{sym}) reduces naturally to the one derived from the HVH theorem \cite{HUGENHOLTZ1958363}, which has been successfully used to calculate symmetry energy of nuclear matter \cite{Xu:2009bb,Xu:2010fh,Xu:2010kf}.

  From Eq. (\ref{sym}), it is easy to find that the nuclear symmetry energy is determined by the product of the nucleon number density and isospin susceptibility. Due to the conservation of baryon numbers, the nucleon number density is directly related to the quark number density with the conditions: $\rho_B=(\rho_u+\rho_d)/3$ and $\rho_I =\rho_d-\rho_u$. Also, the baryon chemical potential and quark chemical potential satisfy the condition: $\mu_I=\frac{\mu_d-\mu_u}{2}$. Thus the nucleon isospin susceptibility equals to the quark isospin susceptibility. This means that the nuclear symmetry energy can be computed in both nucleon and quark degrees of freedom. Therefore, the nuclear symmetry energy can service as an important link between nuclear physics (nucleon degree of freedom) and low energy QCD (quark and gluon degrees of freedom) despite the fact that most calculations of nuclear symmetry energy is performed in the nucleon degree of freedom (see a review in Ref. \cite{Li:2008gp}).

\begin{figure}
\includegraphics[width=0.8\textwidth]{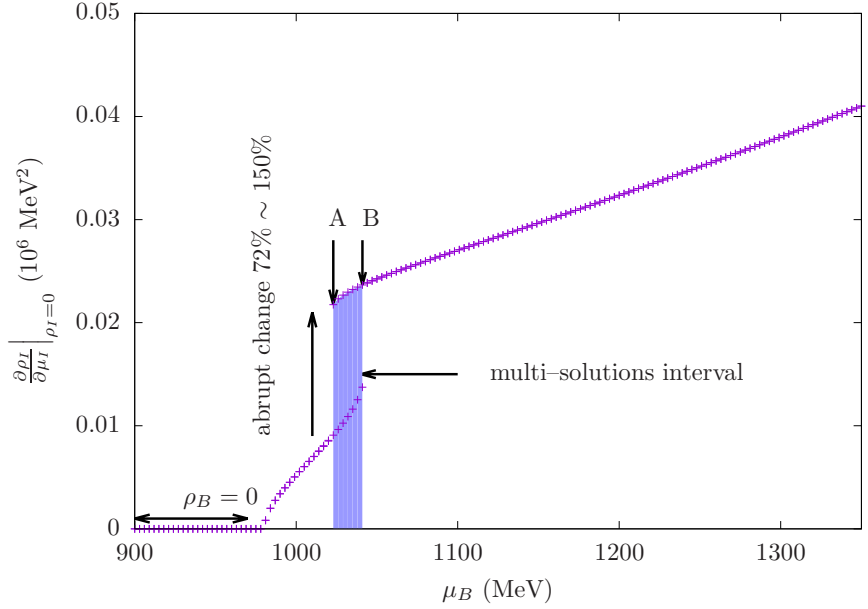}
\caption {(Color online) The isospin susceptibility $\frac{\partial \rho_I}{\partial \mu_I}$ as a function of baryon chemical potential $\mu_B$ at $\rho_I=0$. In the blue shadow interval ($\mu_B$ ranging from 1023 MeV to 1044 MeV), the isospin susceptibility has multi-solutions.}\label{sus}
\end{figure}

More importantly, one can find an interesting result from the nuclear symmetry energy formula Eq. (\ref{sym}), {\it i.e.} the most important term of the symmetry energy formula is the isospin  susceptibility $\frac{\partial \rho_I}{\partial \mu_I}$, which is the linear response of the asymmetry density to the isospin chemical potential. Since the isospin susceptibility couples to the chiral condensate that is considered as the order parameter of chiral phase transition, it can be used to probe the location of chiral phase transition. From the QCD phase diagram, it is found that there undergoes chiral phase transition at zero temperature and finite baryon chemical potential \cite{Ueda:2013sia,Du:2013oza,Cui:2015xta,Lu:2016uwy,Shi:2016koj,Xu:2015vna}, which is considered as a first order phase transition. As is known, the order parameter shows an abrupt change at the critical baryon chemical potential where the  first order phase transition undergoes. To determine the position of chiral phase transition, the Nambu--Jona--Lasinio (NJL) model is adopted to perform calculations. The NJL model is a successful low energy effective theory of QCD, which can capture the main physical features of QCD, for example, the partial restoration of chiral symmetry. The Lagrangian of two flavor NJL model in Minkowski space is given by
\begin{eqnarray}
\mathcal{L}=\bar\psi_f(i\gamma^{\mu}\partial_{\mu}-m)\psi_f+g[(\bar\psi_f\psi_f)^2+(\bar\psi_f i\gamma_5 \tau\psi_f)^2], f=u,d
\end{eqnarray}
where $\psi_f$ denotes two flavor light quarks field. $m$ denotes current quark mass and $g$ denotes coupling constant. In our numerical calculation, the parameters $m$ = 5 MeV, $g$ = 5.02 GeV$^{-2}$ and three--momentum cutoff $\Lambda$ = 653 MeV \cite{Klevansky:1992qe} are adopted and  the obtained isospin  susceptibility as a function of baryon chemical potential from the NJL model  is given in FIG. \ref{sus}. It is shown in FIG. \ref{sus} that, when the baryon chemical potential $\mu_B$ is less than 981 MeV, both the isospin susceptibility and the baryon number density $\rho_B$ are zero since the baryons  can  not be excited from the QCD vacuum within this energy region \cite{Halasz:1998qr}. There exists multi-solutions of the isospin susceptibility in the shadow interval with the baryon chemical potential ranging from 1023 MeV to 1044 MeV.  The critical baryon chemical potential of chiral phase transition is located in this interval (see FIG. \ref{sus}). It is of particular interest that the isospin susceptibility has an abrupt change in the shadow interval. Our numerical calculations show that, the isospin susceptibility is increased by $72\%$ and $150\%$  at the edges {\it A} and {\it B} of the interval, respectively.

\begin{figure}
\includegraphics[width=0.8\textwidth]{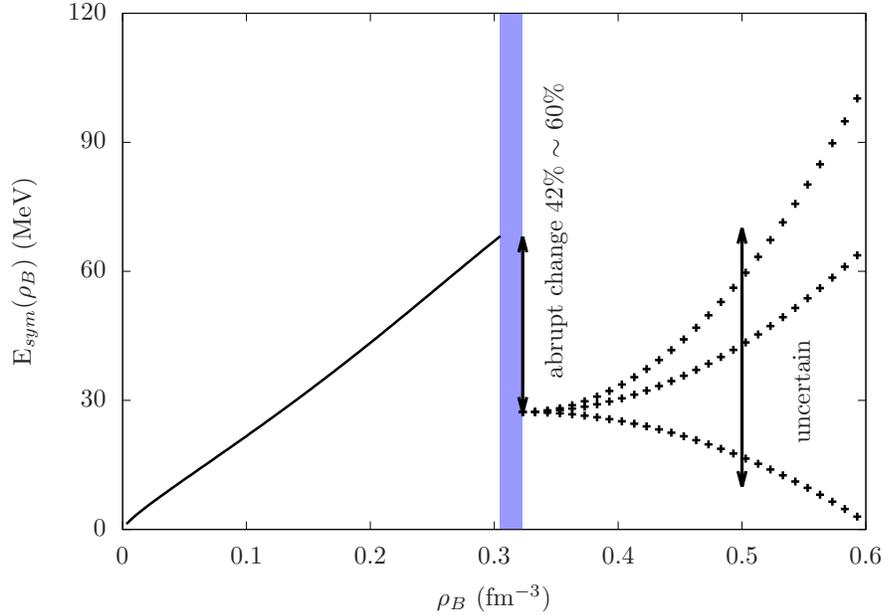}
\caption {(Color online) The nuclear symmetry energy as a function of nuclear matter density $\rho_B$ from the relativistic mean field model with the force parameter NL-B1. The blue interval ($\rho_B$ ranging from 0.305 fm$^{-3}$ to 0.323 fm$^{-3}$) is corresponding to the blue interval ($\mu_B$ ranging from 1023 MeV to 1044 MeV) in FIG. \ref{sus}.} \label{rmf}
\end{figure}

To locate the critical baryon density corresponding to the baryon chemical potential of chiral phase transition, the relativistic mean field (RMF) model (for details, see Ref. \cite{Reinhard:1989zi}) with the force parameter NL-B1 is adopted to perform calculation. It should be noted that it is not possible to pin down the symmetry energy at all densities in present calculations, but the abrupt change of symmetry energy at the critical baryon density can be clearly shown. In Fig. \ref{rmf}, the symmetry energy as a function of the nuclear matter density from the RMF calculations with the NL-B1 force parameter is plotted.
From Fig. \ref{rmf}, it is easy to find that the symmetry energy increases with the nuclear matter density up to $\rho$ = 0.305 fm$^{-3}$ \cite{Chen:2007ih}. The density region corresponding to the interval of the baryon chemical potential in FIG. \ref{sus} ranges from 0.305 fm$^{-3}$ to 0.323 fm$^{-3}$. Because the isospin susceptibility is increased by $72\% \sim 150\%$ at the critical baryon chemical potential, the symmetry energy should be decreased correspondingly at the critical nuclear matter density. As shown by the blue interval in FIG. \ref{rmf}, the symmetry energy is suddenly decreased by $42\% \sim 60\%$ at the density region of the chiral phase transition. When the density is larger than that of the blue interval, the symmetry energy in FIG. \ref{rmf} may have different possibilities as shown by the dotted lines. This sudden decrease of symmetry energy at the critical density of the chiral phase transition could possibly be detected from future nuclear physics experiments.

To summarize, we use the functional path integral method to obtain a model-independent formula for nuclear symmetry energy, which reveals clearly the relation between nuclear symmetry energy and the isospin susceptibility, which can be used to probe the location of chiral phase transition. If the quasi-particle  model is adopted, our formula can reduce naturally to the analytical formula derived from the HVH theorem. Based on our present calculations, it is found that the chiral phase transition is very important in determining the nuclear symmetry energy, which has an abrupt change at the critical density where the chiral phase transition undergoes.

\acknowledgments
This work is supported by the National Natural Science Foundation of China (Grants No.11575082, No.11235001, No.11535004, No.11275097, No.11475085 and No.11535005).

\end{document}